\documentclass[12pt,leqno]{amsart}
\usepackage{amsmath,amsfonts}
\usepackage{amssymb}
\usepackage{amsthm}
\usepackage{hyperref}
\usepackage{graphicx}
\usepackage{bbm}
\usepackage{setspace}
\usepackage{subcaption}
\usepackage{algorithmic}
\usepackage{algorithm}
\usepackage{bm}
\usepackage{rotating}
\usepackage{fancyvrb}
\usepackage{bbm}
\usepackage{scrextend}

\renewcommand{\arraystretch}{1.2}
\newtheorem{theorem}{Theorem}[section]
\newtheorem{lemma}{Lemma}[section]

\renewcommand{\arraystretch}{1.2}
\def\ba{\begin{array}}
\def\ea{\end{array}}
\def\beq{\begin{equation}}
\def\eeq{\end{equation}}
\def\bea{\begin{eqnarray}}
\def\eea{\end{eqnarray}}
\def\beann{\begin{eqnarray*}}
\def\eeann{\end{eqnarray*}}
\def\C{\mathbb{C}}
\def\R{\mathbb{R}}
\def\H{\mathbb{H}}
\def\S{\mathbb{S}}

\def\diag{\textup{diag}}
\def\Diag{\textup{Diag}}

\def\trace{\textup{Tr}}
\def\Null{\textup{null}}
\def\Ra{\textup{range}}

\def\ln{\textup{ln}}

\def\vec{\textup{vec}}
\def\supp{\textup{supp}}

\newcommand{\ignore}[1]{}
\newcommand{\tx}[1]{\texttt{#1}}

\title{\bf Efficient Implementation of Interior-Point Methods for Quantum Relative Entropy}
\author{
Mehdi Karimi \and Levent Tun\c{c}el}
\date{December 12, 2023, (updated: \today)}
 \thanks{\noindent  Mehdi Karimi: Department of Mathematics, Illinois State University, Normal, IL, 61761.  ({e-mail: \bf mkarim3@ilstu.edu}).   Research of this author was supported in part by	 the National
Science Foundation (NSF) under Grant No. CMMI-2347120.\\
 Levent Tun\c{c}el: Department of Combinatorics and Optimization, Faculty of Mathematics,
University
of Waterloo, Waterloo, Ontario N2L 3G1, Canada (e-mail: {\bf 
levent.tuncel@uwaterloo.ca}). Research of this
author was supported in part by Discovery Grants from the Natural
Sciences and Engineering Research
Council (NSERC) of Canada.}

\begin{document}

\begin{abstract}
Quantum Relative Entropy (QRE) programming is a recently popular and challenging class of convex optimization problems with significant applications in quantum computing and quantum information theory. We are interested in modern interior point (IP) methods based on optimal self-concordant barriers for the QRE cone. A range of theoretical and numerical challenges associated with such barrier functions and the QRE cones have hindered the scalability of IP methods. To address these challenges, we propose a series of numerical and linear algebraic techniques and heuristics aimed at enhancing the efficiency of gradient and Hessian computations for the self-concordant barrier function, solving linear systems, and performing matrix-vector products.  We also introduce and deliberate about some interesting concepts related to QRE such as symmetric quantum relative entropy (SQRE). We design a two-phase method for performing facial reduction that can significantly improve the performance of QRE programming. Our new techniques have been implemented in the latest version (DDS 2.2) of the software package DDS. In addition to handling QRE constraints, DDS accepts any combination of several other conic and non-conic convex constraints. Our comprehensive numerical experiments encompass several parts including 1) a comparison of DDS 2.2 with Hypatia for the nearest correlation matrix problem, 2) using DDS 2.2 for combining QRE constraints with various other constraint types, and 3) calculating the key rate for quantum key distribution (QKD) channels and presenting results for several QKD protocols.
\end{abstract}
\maketitle

\pagestyle{myheadings} \thispagestyle{plain}
\markboth{KARIMI and TUN{\c C}EL}
{Interior-Point Methods for Quantum Relative Entropy}

\section{Introduction}
In this manuscript, we propose and evaluate various techniques to efficiently solve convex optimization problems involving the quantum relative entropy (QRE) cone using interior-point methods. Optimization over problems involving the QRE cone has several applications in quantum computing. These applications include calculating the key rate of quantum key distribution (QKD) channels (\cite{scarani2009security,xu2020secure}) or calculating the quantum rate-distortion function (\cite{datta2012quantum,he2023efficient}). QKD is a commercialized secure communication method
that distributes a secret key between two honest parties in the presence of an eavesdropper. The rate-distortion function is a fundamental concept in information theory that quantifies the minimum achievable compression rate for transmitting a source signal within a specified distortion or reconstruction error bound (\cite{cover1999elements}). The quantum relative entropy function is the matrix extension of vector relative entropy or Kullback-Leibler (KL) divergence (\cite{kullback1951information,cover1999elements,van2014renyi,hart2000pattern,chandrasekaran2017relative}) of two vectors which is defined as $KL: \R^n \oplus \R^n \rightarrow \R \cup \{+\infty\}$:
\begin{eqnarray} \label{eq:KL-1}
KL(x,y) := \left\{
\begin{array}{ll}
 \sum_{i=1}^n x_i \ln(x_i) - x_i \ln(y_i),  &  \ \ \ x,y \in \R_+^n, \supp(x) \subseteq \supp(y), \\
 +\infty   &   \ \ \ \text{o.w.}
\end{array}  \right.
\end{eqnarray}
where $\supp(x):=\{i: x_i \neq 0\}$ denotes the support of $x$. 
KL divergence, mostly used to measure the difference of two probability  distributions, is an important function in statistics, information theory, optimization, and machine learning. KL divergence is widely used in machine learning for tasks like information retrieval, clustering, generative modeling, and variational autoencoders (\cite{doersch2016tutorial,goodfellow2016deep}). KL divergence is also used for convergence proof in classic and quantum zero-sum games by \cite{daskalakis2018last,bouland2023quantum}. To define the quantum version of the KL divergence, we need the definition of the matrix extension of a univariate function. 
 Consider a function $f: \mathbb R \rightarrow \mathbb R\cup\{+\infty\}$ and let $X\in \mathbb H^n$ ($\mathbb H^n$ is the set of $n$-by-$n$ Hermitian matrices with complex entries) with a spectral decomposition $X=U \Diag(\lambda_1,\ldots,\lambda_n) U^*$, where $\Diag$ returns a diagonal matrix with the given entries on its diagonal and $U^*$ is the conjugate transpose of a unitary matrix $U$. We define the \emph{matrix extension} $F$ of $f$ as  $F(X) := U \Diag(f(\lambda_1),\ldots,f(\lambda_n)) U^*$. Then, we define the trace of this extension function as 
\begin{eqnarray}\label{eq:fun_cal_1}
\trace(F(X)):= \left\{ \begin{array}{ll}
\trace(U \Diag(f(\lambda_1),\ldots,f(\lambda_n)) U^*)& \text{if $f(\lambda_i) \in \R, \ \forall i$}, \\
+\infty  &  \text{o.w.} 
\end{array} \right.
\end{eqnarray} 
For the special case of $f(x)=x\ln(x)$, we use the convention that $f(0):=0$, so in this special case, $\trace(F(X))$ has a real value for every positive semidefinite matrix. For two matrices $X,Y \in \H_+^n$, the quantity $\trace(X\ln(Y))$ is real if the intersection of the null space of $Y$ and the range of $X$ is the zero vector (equivalently, range of $X$ is contained in the range of $Y$). Then, we can define the quantum relative entropy function $qre: \H^n \oplus \H^n \rightarrow \R \cup \{+\infty\}$ as 
\[
qre(X,Y) :=  \left\{ \begin{array}{ll}
\trace(X\ln(X)-X\ln(Y))& \text{if $X,Y \in \H^n_+$ and $\Ra(X) \cap \Null(Y) = \{0\}$}, \\
+\infty  &  \text{o.w.} 
\end{array} \right.
\] 
The QRE cone is defined as the epigraph of the $qre$ function:
\[
QRE^n:=\left\{(t,X,Y) \in \R \oplus \H_+^n \oplus \H_+^n : qre(X,Y) \leq t \right\}.
\]
QRE programming concerns with optimization problems over the intersection of one or more QRE cones with an affine subspace and potentially many other simpler convex sets. One approach to solve QRE programming is approximating it with other tractable optimization classes such as semidefinite programming (SDP) (\cite{fawzi2019semidefinite,bertsimas2023new}). These SDP approximations are expensive and do not scale well; therefore, work only for small size problems.  We are interested in using modern  interior-point (IP) algorithms for convex optimization based on the theory of self-concordant (s.c.) functions and barriers (\cite{interior-book}). The intriguing theoretical properties exhibited by the s.c.\ barriers for symmetric cones \cite{nesterov1997self,nesterov1998primal}, along with their extensive practical applications, have positioned the primary research emphasis of IP algorithms on symmetric cones. After decades of successful implementation of interior-point algorithms for optimization over symmetric cones (see, for instance, \cite{toh1999sdpt3,sedumi,mosek,DDS}), there have been several recent efforts to create efficient codes for handling other convex sets with available computationally efficient self-concordant (s.c.) barriers. The available modern IP codes for solving convex optimization problems (beyond symmetric cones) using s.c.\ barriers are: a MATLAB-based software package Alfonso (\cite{alfonso}), a software package Hypatia in the Julia language (\cite{coey2022solving}), and a MATLAB-based software package DDS (\cite{DDS}).    DDS has some major differences from the other two, including: 1) accepting both conic and non-conic constraints, 2) utilizing the Legendre-Fenchel conjugate of the s.c.\ barriers when available, and 3) using conservative strategies for the implementation of each iteration, and strict stopping criteria (for improved robustness, see \cite{karimi_status}).

\cite{interior-book} presented a theoretical s.c.\ function (universal barrier) for any closed convex set, which is very costly to compute in general. To apply modern IP methods to optimization problems involving the QRE cone or any other convex set, a computationally efficient s.c.\ barrier in needed, where its gradient and Hessian can be calculated in a reasonable time. DDS has been using the following barrier function since 2019 for solving problems involving quantum relative entropy constraints $\Phi: \R \oplus \H^n \oplus \H^n \rightarrow \R \cup \{+\infty\}$:
\begin{eqnarray} \label{eq:QRE-bar}
\Phi(t,X,Y):= \left\{
\begin{array}{ll}
 -\ln(t - qre(X,Y)) - \ln \det(X) - \ln \det(Y), &  \text{if } X,Y \in \H_{++}^n, \\
 +\infty   &   \text{o.w.}
\end{array}  \right. 
\end{eqnarray}
which was recently proved to be self-concordant by Fawzi and Saunderson (\cite{fawzi2022optimal}). The first available code for QRE programming was CVXQUAD (\cite{cvxquad}), which is a collection of matrix functions to be used on top of CVX (\cite{cvx}). The package CVXQUAD is based on the paper  (\cite{fawzi2019semidefinite}), which approximates the matrix logarithm with functions that can be described by SDPs. CVXQUAD does not scale well and the SDP approximation becomes too large for available SDP solvers even for matrices of size 15. Note that the SDP formulation of the quantum relative entropy function in \cite{fawzi2019semidefinite} involves linear matrix inequalities of size $n^2 \times n^2$. \cite{faybusovich2020self} proved self-concordance results for some functions related to quantum entropy and then \cite{fay-QRE} designed some interior-point algorithms for  various problems involving quantum entropy. They also consider minimizing the $qre(X,Y)$ function for the special case of calculating the key rate for quantum key distribution (QKD) channels, which is one of the most popular applications of QRE programming. For this special case, both $X$ and $Y$ are linear functions of another matrix $\rho$ and the problem formulation is significantly simpler than our general QRE setup\footnote{The code of \cite{fay-QRE} is not publicly available.}. \cite{hu2022robust} also proposed an interior-point method, not using the s.c.\ barriers for the QRE cone, for solving QKD key rate using the simplified formulation. As far as we know, Hypatia and DDS are the only publicly available codes for QRE programming where both use the s.c.\ barrier in \eqref{eq:QRE-bar}. The main differences of DDS (and Hypatia) with \cite{fay-QRE} and \cite{hu2022robust} are:
\begin{itemize}
\item DDS QRE programming is not just for the simplified QKD key rate computation, but is for any problem with a combination of an arbitrary number of QRE cones, with all the other function/set constraints available in DDS. 
\item DDS is using the optimal s.c.\ barrier in \eqref{eq:QRE-bar}. \cite{fay-QRE} use their conjectured s.c.\ function for the simplified problem, and \cite{hu2022robust} do not utilize self-concordance. 
\end{itemize}

Several computational and theoretical challenges related to \eqref{eq:QRE-bar} have hindered the scalability of Quantum Relative Entropy (QRE) optimization solvers, specifically DDS 2.1 and Hypatia.   Among the primary issues are the complexity of evaluating the gradient and Hessian of $\Phi$ in \eqref{eq:QRE-bar}, and also solving the linear systems involving the Hessian, which are needed in implementing the second-order IP methods. In this paper, we present a set of numerical and theoretical techniques aimed at enhancing the performance of IP methods, and then evaluate the effectiveness of these techniques through a series of numerical experiments. The new techniques have been implemented in DDS 2.2 (\cite{DDS2.2}), which is introduced and released by this paper. Here are the contributions of this paper:
\begin{itemize}
\item Introducing numerical and linear algebraic techniques and heuristics to improve the calculation of the gradient and Hessian of $\Phi$, solving the needed linear systems, and calculating the matrix-vector products. These techniques improved DDS 2.2, and enabled us to solve much larger instances compared to DDS 2.1 and Hypatia (Sections \ref{sec:2} and \ref{sec:3}).
\item Introducing and deliberating about the concept of \emph{symmetric quantum relative entropy} (Section \ref{sec:SQRE}).
\item Introducing a two-phase approach for QRE programming to improve the running time and condition of the problems (Section \ref{sec:2phase}).  
\item Developing a comprehensive setup (including a two-phase approach and facial reduction) for calculating quantum key distribution (QKD) channel rates (Section \ref{sec:QKD}), and discussing how to handle complex Hermitian matrices (Section \ref{sec:4}).
\item A comprehensive numerical experiment including: 1) comparison of DDS 2.2 with Hypatia for the nearest correlation matrix, 2) using DDS for combination of QRE and many other types of constraints, and 3) examples to elaborate on the two-phase method and its performance improvement, 4) Solving symmetric QRE programming problems, 5) calculating the key rate for QKD channels and presenting results for several QKD protocols (Section \ref{sec:num}).  
\end{itemize}

\subsection{Notations} The sets $\S^n$, $\S^n_+$, and $\S^n_{++}$ are the set of $n$-by-$n$ symmetric matrices, positive semidefinite matrices, and positive definite matrices, respectively. For a multivariate function $f$, both $f'$ and $\nabla f$ are used for the gradient, and both $f''$ and $\nabla^2 f$ are used for the Hessian. 

\section{Evaluating the Derivatives for Quantum Relative Entropy} \label{sec:2} 
In this section, we discuss how to calculate the gradient and Hessian for the s.c.\ barrier function in \eqref{eq:QRE-bar}. The details of our Domain-Driven infeasible-start predictor-corrector algorithms are given in \cite{karimi2020primal,karimi_status}. At each iteration of the algorithm, for both the predictor and corrector steps, we solve a linear system of the form:
\begin{eqnarray} \label{eq:system}
\left( U^\top  \underbrace{ \left[ \begin{array}{cc} \bar H  &  0  \\  0  &  \left [\hat H \right] ^{-1} \end{array} \right]}_{\mathcal H(\bar H, \hat H)} U \right)   d=r_{RHS}, 
\end{eqnarray}
where $\bar H$ and $\hat H$ are positive definite matrices based on the Hessians of the s.c.\ barrier $\Phi$ representing the feasible set, $U$ is a fixed matrix, and $r_{RHS}$ is the right-hand-side vector that is different for the predictor and corrector steps. The following pseudocode gives a framework for the algorithm (\cite{karimi2020primal,karimi_status}):

\noindent\rule{16cm}{0.6pt}\\
\noindent \textbf{Framework for the Interior-Point Algorithm in DDS} \vspace{-0.3cm} \\
\noindent\rule{16cm}{0.6pt}\\
\noindent {\bf INPUT:} Matrix $U$, the oracles for calculating $\Phi$ and its derivatives. A proximity measure $\Omega$ and the constants $0 < \delta_1 < \delta_2$.   \\
\noindent {\bf while} (the stopping criteria are not met) 
\begin{addmargin}{1cm}
 {\bf Predictor Step} 
\end{addmargin} 
\begin{addmargin}{1cm}
Calculate the predictor search direction $d$ by solving \eqref{eq:system} with the proper $r_{RHS}$. Update the current point using the search direction and a step size that guarantees  $\Omega  \leq \delta_2$.
\end{addmargin} 
\begin{addmargin}{1cm}
{\bf Corrector Step} 
\end{addmargin}  
\begin{addmargin}{1cm}
Modify the system and the $r_{RHS}$ in \eqref{eq:system} and calculate the corrector search direction $d$. Apply the corrector step at least once so that the updated point satisfies $\Omega < \delta_1$. 
\end{addmargin}
\begin{addmargin}{1cm}
{\bf Fine Tuning} 
\end{addmargin}  
\begin{addmargin}{1cm}
Modify the dual iterate to make sure it approximately satisfies dual feasibility.
\end{addmargin}
\noindent {\bf end while}\\
\noindent\rule{16cm}{0.6pt}\\

For QRE programming, the main challenge in forming the linear system  \eqref{eq:system} is calculating the gradient and Hessian for the $qre(X,Y)$ function in an efficient and numerically stable way. For this, we need to calculate the derivatives for $\trace(X\ln(X))$ and $\trace(-X\ln(Y))$. $X\ln(X)$ is the matrix extension of $x\ln(x)$. For the trace of a general matrix extension function $\trace(F(X))$ defined in \eqref{eq:fun_cal_1}, the gradient can be calculated by the following theorem:
\begin{theorem} [see, for example, \cite{hiai2014introduction}-Section 3.3] \label{thm:mtx_fun_der_1}
Let $X$ and $H$ be self-adjoint matrices and $f: (a,b)\mapsto \mathbb R$ be a continuously differentiable function defined on an interval. Assume that the eigenvalues of $X+\alpha H$ are in $(a,b)$ for an interval around $\alpha_0 \in \mathbb R$. Then, 
\begin{eqnarray}\label{eq:fun_cal_5}
\left. \frac{d}{d\alpha}\trace{F(X+\alpha H)}\right|_{\alpha=\alpha_0} = \trace{HF'(X+\alpha_0H)}. 
\end{eqnarray}
\end{theorem}
By putting $\alpha_0=0$ in \eqref{eq:fun_cal_5}, we get the directional derivative of $F$ in the direction of $H$, and we can write:
\begin{eqnarray*} \label{eq:QRE-9-2}
\left. \frac{d}{d\alpha}\trace{F(X+\alpha H)}\right|_{\alpha=0} = \trace{HF'(X)} = \vec(F'(X))^\top \vec(H), 
\end{eqnarray*}
where $\vec$ changes a matrix into a vector by stacking the columns on top of one another. 
This implies that if we look at $(\trace(F(X)))'$ as a vector of length $n^2$, we have 
\begin{eqnarray} \label{eq:QRE-9}
(\trace(F(X)))' = \vec(F'(X)).
\end{eqnarray}
For the rest of our discussion, we need two definitions for univariate functions similar to the derivative. For a  continuously differentiable function $f:(a,b) \mapsto \R$, we define the first and second divided differences as 
\begin{eqnarray} \label{eq:div-dif}
f^{[1]}(\alpha,\beta) &:=& \left\{\begin{array}{ll} \frac{f(\alpha)-f(\beta)}{\alpha-\beta} & \alpha \neq \beta, \\ f'(\alpha) & \alpha=\beta. \end{array} \right.  \nonumber \\
f^{[2]}(\alpha,\beta,\gamma) &:=& \left\{\begin{array}{ll}  \frac{f^{[1]}(\alpha,\beta)-f^{[1]}(\alpha,\gamma)}{\beta-\gamma} &  \beta \neq \gamma,  \\ \frac{f^{[1]}(\alpha,\beta)-f'(\beta)}{\alpha-\beta} & \beta = \gamma \neq \alpha, \\ -\frac{1}{2} f''(\alpha) &   \beta = \gamma = \alpha. \end{array} \right. 
\end{eqnarray}

To calculate the Hessian of $\trace(F(X))$, we can use the following theorem: 
\begin{theorem} [see, for example, \cite{hiai2014introduction}-Theorem 3.25] \label{thm:mtx_fun_der_2}
Assume that $f:(a,b) \mapsto \R$ is a $\mathcal C^1$-function and $T=\Diag(t_1,\ldots,t_n)$ with $t_i \in (a,b), \ i\in\{1,\ldots,n\}$. Then, for a Hermitian matrix $H$, we have
\begin{eqnarray} \label{eq:thm:mtx_fun_der_2_1}
\left. \frac{d}{d\alpha} F(T+\alpha H) \right|_{\alpha=0}=T_f \odot H, 
\end{eqnarray}
where $\odot$ is the Hadamard product and $T_f$ is the divided difference matrix defined as
\begin{eqnarray} \label{eq:thm:mtx_fun_der_2_2}
[T_f]_{ij}:= f^{[1]}(t_i,t_j), \ \ \ \forall i,j \in \{1,\ldots,n\}. 
\end{eqnarray}
\end{theorem}
$T$ is diagonal in the statement of the theorem, which is without loss of generality. Note that by the definition of functional calculus in \eqref{eq:fun_cal_1}, for a Hermitian matrix $X$ and a unitary matrix $U$, we have
\begin{eqnarray}\label{eq:fun_cal_13}
F(UXU^*)=UF(X)U^*. 
\end{eqnarray}
Therefore,  for a matrix $T=U\Diag(t_1,\ldots,t_n)U^*$, we can update \eqref{eq:thm:mtx_fun_der_2_1} as
\begin{eqnarray} \label{eq:thm:mtx_fun_der_2_3}
\left. \frac{d}{d \alpha} F(T+\alpha H) \right|_{\alpha=0}=U \left(T_f \odot (U^*HU)\right) U^*,
\end{eqnarray}
where we extend the definition of $T_f$ in \eqref{eq:thm:mtx_fun_der_2_2} to non-diagonal matrices by defining that $T_f:= (U^* T U)_f$. In other words, for a non-diagonal matrix $T=U\Diag(t_1,\ldots,t_n)U^*$, we use $\Diag(t_1,\ldots,t_n)$ to calculate $T_f$ using \eqref{eq:thm:mtx_fun_der_2_2}.
Now we can use Theorems  \ref{thm:mtx_fun_der_2} and \ref{thm:mtx_fun_der_1} to calculate the Hessian of the function $\trace(F(X))$. 
\begin{theorem}
Let $X$, $H$, and $\tilde H$ be self-adjoint matrices and $f: (a,b)\mapsto \mathbb R$ be a continuously differentiable function defined on an interval. Assume that the eigenvalues of $X+tH$ and $X+t \tilde H$ are in $(a,b)$ for an interval around $t=0$. Assume that $X=U_X\Diag(\lambda_1,\ldots,\lambda_n)U_X^*$. Then, 
\begin{eqnarray}\label{eq:fun_cal_14}
\nabla^2 \trace(F(X))[H,\tilde H]=\trace \left( \left(X_{f'} \odot (U_X^*HU_X)\right) U_X^* \tilde HU_X \right). 
\end{eqnarray}
\end{theorem}
\proof{Proof.}
We can write
\begin{eqnarray} \label{eq:proof-1}
\begin{array}{rlc}
\nabla^2 \trace(F(X))[H,\tilde H] =  &  \left.\frac{d}{d\beta} \left.\frac{d}{d\alpha} \trace(F(X+\beta H + \alpha \tilde H)) \right|_{\alpha=0} \right|_{\beta=0} & \\ 
  =& \left.\frac{d}{d\beta} \trace(H F'(X+\beta  H)) \right|_{\beta=0},  & \text{using \eqref{eq:fun_cal_5}} \\
  =& \trace(\tilde H\left.\frac{d}{d\beta}  F'(X+\beta  H) \right|_{\beta=0})  & \\
  =& \trace(\tilde H U_X \left(X_{f'} \odot (U_X^*HU_X)\right) U_X^*),  & \text{using \eqref{eq:thm:mtx_fun_der_2_3}}\\
  =& \trace \left( \left(X_{f'} \odot (U_X^*HU_X)\right) U_X^* \tilde HU_X \right).
\end{array}
\end{eqnarray}
  \endproof
To find a formula for the matrix form of the Hessian, note that by using the properties of the Hadamard product, we have
\[
\vec(X_{f'} \odot (U_X^*HU_X)) = \Diag(\vec(X_{f'}))  \vec(U_X^*HU_X).
\]
Using this, we have
\begin{eqnarray} \label{eq:proof-2}
\begin{array}{rl}
\trace \left( \left(X_{f'} \odot (U_X^*HU_X)\right) U_X^* \tilde HU_X \right) = & \vec(X_{f'} \odot (U_X^*HU_X))^\top \vec(U_X^* \tilde HU_X) \\
   =& \vec(U_X^*HU_X)^\top  \Diag(\vec(X_{f'})) \vec(U_X^* \tilde HU_X)  \\
   =&  \vec(H)^\top  (U_X \otimes U_X) \Diag(\vec(X_{f'})) (U_X^* \otimes U_X^*) \vec(\tilde H).
\end{array}
\end{eqnarray}
So the matrix form of the Hessian is
\begin{eqnarray} \label{eq:QRE-10}
(\trace(F))''(X) = (U_X \otimes U_X) \Diag(\vec(X_{f'})) (U_X^* \otimes U_X^*).
\end{eqnarray}

For the other components of the Hessian of $qre$, we need to differentiate $\trace(-X\ln(Y))$ in terms of $X$ and $Y$. In terms of $Y$, for a fixed matrix $X$ and a continuously differentiable function $f: (a,b)\mapsto \mathbb R$, let us define 
\begin{eqnarray} \label{eq:QRE-5}
F_X(Y):= \trace(X F(Y)).
\end{eqnarray}
Let $Y=U_Y \Diag(\gamma_1,\ldots,\gamma_n) U_Y^*$ be the spectral decomposition of $Y$. The gradient of $F_X(Y)$  can be calculated using Theorem \ref{thm:mtx_fun_der_2} as:
\begin{eqnarray} \label{eq:QRE-6}
F'_X(Y) = U_Y \left(Y_f \odot (U_Y^*XU_Y)\right) U_Y^*.
\end{eqnarray}
The Hessian of $F_X(Y)$ is calculated as follows in \cite{fay-QRE}:
\begin{eqnarray} \label{eq:QRE-7}
F''_X(Y) = (U_Y \otimes U_Y) S (U_Y^* \otimes U_Y^*),
\end{eqnarray}
where $S$ is the $n^2$-by-$n^2$ second divided difference matrix. If we assume that $S$ is a block matrix of size $n$-by-$n$ where each block is again a matrix of size $n$-by-$n$,
then we can show the entries of $S$ as $S_{ij,kl}$ where $ij$ denotes the place of the block, and $kl$ denotes the rows and columns inside the block. We have:
\begin{eqnarray} \label{eq:QRE-8}
S_{ij,kl} = \delta_{kl} \tilde X_{ij} f^{[2]}(\gamma_i,\gamma_j,\gamma_l) + \delta_{ij} \tilde X_{kl} f^{[2]}(\gamma_j,\gamma_k,\gamma_l),
\end{eqnarray}
where $\tilde X := U_Y^* X U_Y$ and $\delta_{ij}$ is an indicator function which is 1 if $i=j$, and 0 otherwise. 
Putting together all these results, we have
\[
qre''(X,Y) = \left[\begin{array}{cc}
H_{11}  & H_{12} \\
H_{12}^\top & H_{22} 
\end{array}\right],
\]
\[ 
\begin{array}{l}
H_{11} = (U_X \otimes U_X) (\Diag(\vec(X_{\ln}))) (U_X \otimes U_X),\\ [1em]
H_{12} =  -(U_Y \otimes U_Y) (\Diag(\vec(Y_{\ln}))) (U_Y \otimes U_Y),\\ [1em]
H_{22} = -(U_Y \otimes U_Y) S (U_Y^* \otimes U_Y^*).
\end{array}
\]
By having the derivatives of the $qre$ function, we can calculate the derivatives for the s.c.\ barrier function $\Phi$ in \eqref{eq:QRE-bar}. For simplicity, we define $T:=t-qre(X,Y)$. We have 
\begin{eqnarray} \label{eq:QRE-11}
\Phi'(t,X,Y) = \left [\begin{array}{c}
\frac{-1}{T} \\  \frac{1}{T} h +\vec(- X^{-1}) \\
\frac{1}{T} \bar h + \vec(- Y^{-1})
\end{array}  \right], \ \ 
\begin{array}{l}
h:= \vec(I+\ln(X)-\ln(Y)) \\
\bar h:= \vec(\left(U_Y \left(Y_f \odot (U_Y^*XU_Y)\right) U_Y^* \right)).
\end{array}
\end{eqnarray}
We can write the Hessian as:
\begin{eqnarray} \label{eq:QRE-12}
&&\Phi''(t,X,Y) = \nonumber \\ && \underbrace{\left [\begin{array}{ccc}
\frac{1}{T^2}  &  -\frac{1}{T^2} h^\top   &  \frac{1}{T^2} \bar h^\top \\  
-\frac{1}{T^2} h  &  \frac{1}{T} H_{11} + (X^{-1} \otimes X^{-1})  &  \frac{1}{T} H_{12} \\
\frac{1}{T^2} \bar h  & \frac{1}{T} H_{12}^\top &  \frac{1}{T} H_{22} + (Y^{-1} \otimes Y^{-1})
\end{array}  \right]}_{\bar H}+\left [\begin{array}{c} 0 \\ \frac{1}{T} h \\ \frac{1}{T} \bar h \end{array} \right] \left [\begin{array}{c} 0 \\ \frac{1}{T} h \\ \frac{1}{T} \bar h \end{array} \right]^\top.
\end{eqnarray}
\subsection{Other numerical techniques}

For calculating the gradient and Hessian of $qre(X,Y)$, numerical instability happens in calculating $X_{\ln}$ and $Y_{\ln}$, where for a matrix $X=U_X \Diag(\lambda_1,\ldots,\lambda_n) U_X^*$, using \eqref{eq:thm:mtx_fun_der_2_2}, we have: 
\begin{eqnarray*}
[X_{\ln}]_{ij} = \ln^{[1]}(\lambda_i,\lambda_j) &=& \left\{\begin{array}{ll} \frac{1}{\lambda_i} & \lambda_i=\lambda_j \\[.5em]
 \frac{\ln(\lambda_i)-\ln(\lambda_j)}{\lambda_i-\lambda_j} & \lambda_i \neq \lambda_j. \end{array} \right.
\end{eqnarray*}
To make the calculation more stable, \cite{faybusovich2020self} used the following equivalent formula given in \cite{higham2008functions}:
\begin{eqnarray*}
\ln^{[1]}(\lambda_i,\lambda_j) &=& \left\{\begin{array}{ll} \frac{1}{\lambda_i} & \lambda_i=\lambda_j \\ [.5em]
\frac{\ln(\lambda_i)-\ln(\lambda_j)}{\lambda_i-\lambda_j} & \lambda_i < \frac{\lambda_j}{2} \ \ \text{or} \ \  \lambda_j < \frac{\lambda_i}{2}  \\[.5em]
\frac{2 \textup{tanh}^{-1}(z)}{\lambda_i-\lambda_j}  & \textup{o.w.} \end{array} \right. 
\end{eqnarray*}
where $z = (\lambda_i-\lambda_j)/(\lambda_i+\lambda_j)$. Numerical experiments with DDS 2.2 shows that this formula indeed works better in terms of numerical stability.  
\section{Solving the Linear System } \label{sec:3} 
In QRE programming with DDS, other than solving the linear system \eqref{eq:system}, for calculating the proximity measure $\Omega$ and the fine-tuning step, we need to solve linear systems that directly have the Hessian on the left-hand-side. In this section, we discuss how to solve these systems in DDS to significantly speed up the algorithm. 
Writing the Hessian as in \eqref{eq:QRE-12} is efficient since the Hessian is the summation of a simpler matrix and a rank one matrix, and we can use Sherman–Morrison formula\footnote{Sherman–Morrison formula has been employed in solving convex optimization problems for a long time. These algorithms go back to the implementation of quasi-Newton methods, implementation of ellipsoid methods (see \cite{BGT1981} and references therein), as well as interior-point methods (see \cite{Karmarkar1984}, \cite{TV2023} and references therein)} to solve the linear systems involving the Hessian. By the Sherman–Morrison formula, the linear system reduces to a linear system with $\bar H$ defined in \eqref{eq:QRE-12}. With some linear algebra, the main part of the reduced linear system is the one involving the matrix
\begin{eqnarray} \label{eq:QER-25}
\left [\begin{array}{cc}
  \frac{1}{T} H_{11} + (X^{-1} \otimes X^{-1})  &  \frac{1}{T} H_{12} \\
 \frac{1}{T} H_{12}^\top &  \frac{1}{T} H_{22} + (Y^{-1} \otimes Y^{-1})
\end{array}  \right].
\end{eqnarray}
Calculating $U_X \otimes U_X$ and $U_Y \otimes U_Y$ is the main computational challenge in forming this matrix. $U_X$ and $U_Y$ are dense matrices, but have the nice property that both are unitary matrices, which we can exploit.  By defining $1/\lambda := \Diag(1/\lambda_1,\ldots,1/\lambda_n)$ and $1/\gamma := \Diag(1/\gamma_1,\ldots,1/\gamma_n)$, we can write the diagonal block matrices of \eqref{eq:QER-25} as
\begin{eqnarray*}
\frac{1}{T} H_{11} + (X^{-1} \otimes X^{-1}) &=& (U_X \otimes U_X) ( \Diag(\frac{1}{T}\vec(X_{\ln})+1/\lambda \otimes 1/\lambda))(U_X \otimes U_X) \\
\frac{1}{T} H_{22} + (Y^{-1} \otimes Y^{-1}) &=& (U_Y \otimes U_Y) (-\frac{1}{T} S+\Diag(1/\gamma \otimes 1/\gamma)) (U_Y^* \otimes U_Y^*).
\end{eqnarray*}

The approximation we made in DDS 2.2 for solving the linear systems with the Hessian of $\Phi$ on the left hand side is  ignoring the off-diagonal block matrices in \eqref{eq:QER-25}. Doing this, the matrix can be factorized as
\begin{eqnarray*}
&& U_{XY}
\left [\begin{array}{cc}\Diag(\frac{1}{T}\vec(X_{\ln})+1/\lambda \otimes 1/\lambda)  & 0 \\ 0 &  -\frac{1}{T} S+\Diag(1/\gamma \otimes 1/\gamma) \end{array} \right]
U_{XY}^*, \\
&& U_{XY} := \left [\begin{array}{cc}U_X \otimes U_X & 0 \\ 0 &  U_Y \otimes U_Y \end{array}  \right].
\end{eqnarray*}
By this simplification, solving the linear system is reduced to solving a system with the matrix in the middle and mostly a system involving $S$. Note that in DDS 2.2, this simplification is not used for the linear system \eqref{eq:system} for calculating the search directions of the IP method. For those systems, the matrix on the left hand side is a quadratic function of $U$, and the Hessian does not directly appear in the left-hand-side. 
\section{Symmetric Quantum Relative Entropy} \label{sec:SQRE}
As vector relative entropy or Kullback-Leibler divergence is not symmetric, there is a natural way to symmetrize the KL function defined in \eqref{eq:KL-1} as $J: \R^n \oplus \R^n \rightarrow \R \cup \{+\infty\}$:
\begin{eqnarray} \label{eq:S1}
J(x,y) := \left\{
\begin{array}{ll}
 KL(x,y) + KL(y,x),  & \ \ \  x,y \in \R_+^n, \supp(x) = \supp(y), \\
 +\infty   &   \ \ \ \text{o.w.}
\end{array}  \right.
\end{eqnarray}
$J(x,y)$ is called  Jeffreys divergence or symmetrized Kullback-Leibler divergence (\cite{kullback1951information,jeffreys1998theory}). In a similar way, we define the symmetric QRE function as $sqre:  \H^n \oplus \H^n \rightarrow \R \cup \{\infty\}$:
\begin{eqnarray} \label{eq:S2}
\ \ \ \ \ \ \ sqre(X,Y) :=  \left\{ \begin{array}{ll}
qre(X,Y)+qre(Y,X) & \ \ \ \text{$X,Y \in \H^n_+$,  $\Ra(X) = \Ra(Y)$}, \\
+\infty  &  \ \ \ \text{o.w.} 
\end{array} \right.
\end{eqnarray}
SQRE is a straightforward extension of Jeffreys divergence, and it was also suggested in the context of QRE (\cite{stone}). Clearly $sqre$ is also a convex function in $(X,Y)$. A code that accepts QRE constraints can also handle SQRE constraints using the following reformulation:
\begin{eqnarray} \label{eq:S3}
qre(X,Y) + qre(Y,X)  \leq t  \ \   \equiv  \left \{ \ \ \begin{array}{rcl}
t_1 + t_2 &\leq& t \\
qre(X,Y)  & \leq & t_1 \\
qre(Y,X)  & \leq & t_2.
\end{array} \right.
\end{eqnarray}
The only issue with this approach is that by this reformulation, the s.c.\ barrier assigned to two QRE constraints has parameter $2n$. However, it is plausible that there exists an efficient s.c.\ barrier with a better parameter for the epigraph of $sqre$. In Subsection \ref{subsec:SQRE}, we present some numerical results to show that formulation \eqref{eq:S3} works well in practice. 
\section{Two-phase Methods} \label{sec:2phase}
In this section, we propose two-phase methods for solving QRE programming which can significantly improve the running time and condition of the problem. This two-phase approach is different than, for example, the one for solving LP problems or SDP problems where the two phases are roughly equivalent in terms of size and complexity. Here, phase one for the QRE programming is a convex optimization problem that can be solved more efficiently and much faster compared to the QRE problem. The solution of Phase-I is used to reformulate the QRE problem to make it a smaller size and well-conditioned (or, at least, better conditioned) QRE programming problem. 

\noindent\rule{16cm}{0.6pt}\\
\noindent \textbf{Framework for the Two-Phase QRE Programming} \vspace{-0.3cm} \\
\noindent\rule{16cm}{0.6pt}\\
\noindent {\bf INPUT:} A QRE optimization problem $(P_{QRE})$.    \\
\noindent {\bf Phase-I:} Create an auxiliary optimization problem $(P_{AUX})$, which is more robust and much faster to solve by DDS.  \\
\noindent {\bf Reformulation:} Use the solution of $(P_{AUX})$ to reformulate $(P_{QRE})$ as  a smaller size and  better conditioned QRE programming problem $(\bar P_{QRE})$.   \\
\noindent {\bf Phase-II:} Use DDS to solve $(\bar P_{QRE})$.  \\
\noindent {\bf Solution Reconstruction:} Use the solution of $(\bar P_{QRE})$ to create a corresponding solution for $(P_{QRE})$.  \\
\noindent\rule{16cm}{0.6pt}\\
Consider an optimization problem of the form
\begin{eqnarray} \label{eq:QRE-28}
\begin{array}{ccll}
\textup{min} &  & qre\left(\sum_{i=1}^k x_i A_i, M \right) & \\
 & & \ell \leq x \leq u, &  \\
\end{array}
\end{eqnarray}
where $M \in \S_+^n$ is given. Assume that the set of points $\sum_{i=1}^k x_i A_i \in \S_+^n$ lie on a smaller \emph{face} of the $\S_+^n$ cone (see, for instance, Chapter 2 of \cite{tuncel-book}). In other words, there exist a positive integer $r < n$ and an $n$-by-$r$ matrix $V$ with orthonormal columns such that 
\[
\left\{X : X=\sum_{i=1}^k x_i A_i \right\} \cap \S_+^n  \subset V \S_+^r V^\top. 
\]
We have the freedom to model the phase-I problem to not just calculate a matrix $V$, but also calculate some other useful information. We present two methods for problem \eqref{eq:QRE-28}: 
\subsection{Dual Method} \label{subsec:5-1} We can find a facial reduction matrix $V$ by solving the following optimization problem 
\begin{eqnarray} \label{eq:QRE-29}
\begin{array}{ccll}
\text{(Phase-I)} & \textup{inf} & -\ln(\det(Y)) & \\
 & &\langle Y, A_i \rangle = 0,&  i \in \{1,\ldots,k\}.  \\
\end{array}
\end{eqnarray}
In the above, $-\ln(\det): \S^n \rightarrow \R \cup \{+\infty\}$ is defined as
\begin{eqnarray*}
-\ln(\det(Y)) := \left\{
\begin{array}{ll}
 -\sum_{j=1}^n \ln(\lambda_j(Y)),  &  \text{if } Y \in \S_{++}^n, \\
 +\infty   &   \text{o.w.}
\end{array}  \right.
\end{eqnarray*}
Indeed, problem \eqref{eq:QRE-29} may be unbounded and generally may not have an optimal solution. However, whenever \eqref{eq:QRE-29} has a nontrivial feasible solution $Y$, we expect a robust interior-point algorithm to generate iterates $Y^{(l)} \in  \S^n_{++}$, approximately satisfying the  equations in \eqref{eq:QRE-29} such that $Y^{(l)}$ approach maximum rank positive semidefinite solutions of \eqref{eq:QRE-29}. Therefore, using such  $Y^{(l)}$ we can approximately identify the minimal face of $\S^n_+$ which contains all optimal solutions of QRE problem \eqref{eq:QRE-28}. 

Let $Y^* \in \S^n_+$ be a solution of \eqref{eq:QRE-29}, then we can show that the columns of $V$ can be chosen as an orthonormal basis for the null space of $Y^*$, since we have 
\begin{eqnarray} \label{eq:QRE-29-2}
Y^* \left(\sum_{i=1}^k x_i A_i\right) = 0,   \ \ \forall x.
\end{eqnarray}
Note however that due to the fact that strict complementarity can fail for SDPs, the complementary face identified by $Y^*$ may be a strict super-set of the minimal face for \eqref{eq:QRE-28} (and one might need to do a full facial reduction to obtain a description of the minimal face for \eqref{eq:QRE-28} which in the worst case may require $(n-1)$ recursive applications of the above process, see a family of examples in \cite{tuncel-book}-Page 43). In our experiments we only performed the above described facial reduction step once. 

In defining problem \eqref{eq:QRE-29}, we ignored the primal constraints $ \ell \leq x \leq u$. In some applications, it might be worthwhile to include such constraints in deriving the corresponding Phase-I problem such that the underlying cone is not $\S_+^n$ but $\S_+^n \oplus \R_+^k \oplus \R_+^k$.
\subsection{Primal Method} We can use the following phase-I problem with the same feasible region as the QRE problem:
\begin{eqnarray} \label{eq:QRE-28-2}
\begin{array}{ccll}
\text{(Phase-I)}  &\textup{inf}   & -\ln(\det(\sum_{i=1}^k x_i A_i)) & \\
 & & \ell \leq x \leq u. &  \\
\end{array}
\end{eqnarray}
As we mentioned in Subsection \ref{subsec:5-1}, here too, a robust interior-point algorithm generates a sequence of vectors $x^{(l)}$ such that $\ell \leq x^{(l)} \leq u$, $\sum_{i=1}^k x^{(l)}_i A_i \in \S_{++}^n $, and $\sum_{i=1}^k x^{(l)}_i A_i$ approach the relative interior of the minimal face of $\S_+^n$ containing all optimal solutions of the QRE problem \eqref{eq:QRE-28}. 
Additionally, solving \eqref{eq:QRE-28-2} gives us a feasible solution that can be given to the QRE solver as an initial point. 

One important feature of Phase-I problems \eqref{eq:QRE-29} and \eqref{eq:QRE-28-2} is that both problems are minimizing a s.c.\ barrier, which typically is a very well-behaved convex optimization problem with local quadratic convergence (\cite{interior-book}). Additionally, both problems can be reformulated as SDPs, which are still more robust and faster compared to the QRE problem. Since the current version of DDS does not support minimizing a s.c.\ barrier (although the ingredients of such an algorithm are already present in the DDS code), we use the SDP reformulations for the numerical results in Subsection \ref{subsec:2p}. The numerical results confirm that the two-phase approach can significantly improve the conditioning and speed of QRE problems. 

In the following section, we propose similar two-phase methods designed for calculating the rate of QKD channels. In the numerical result section, we show that two-phase methods can significantly improve the size and condition of the QRE problems, in general and also in the context of QKD channel rate calculations. 
\section{Quantum Key Distribution Rate}  \label{sec:QKD}
One application of minimizing $qre$ function is calculating the rate of quantum key distribution (QKD) channels. QKD is a secure communication method between two parties involving components of quantum mechanics (\cite{QKD1}). The security of the QKD channel depends on the exact calculation of its key rate. There are different protocols for QKD and for many of them, the main non-trivial component of calculating the key rate is an optimization problem of the form:
\begin{eqnarray}  \label{eq:num-3}
&\min& qre(\mathcal G(\rho), \mathcal Z(\mathcal G(\rho))) \nonumber \\
& \text{s.t.}& A(\rho) = b , \nonumber  \\
&  & \rho \succeq 0,
\end{eqnarray}
where $A$ is a linear map on Hermitian matrices and $\mathcal G$ and $\mathcal Z$ are Kraus operators. The Linear map $\mathcal G: \mathbb H^n \rightarrow \mathbb H^k$ is defined as
\begin{eqnarray}
\mathcal G(\rho) := \sum_{j=1}^{n_g} K_j \rho K_j^\dag,
\end{eqnarray}
where $K_j \in \C^{k \times n}$ and $\sum_{j=1}^{n_g} K_j K_j^\dag \preceq I$, and the self-adjoint linear map $\mathcal Z: \H^k \rightarrow \H^k$ is defined as
\begin{eqnarray}
\mathcal Z(\delta) := \sum_{j=1}^{n_z} Z_j \delta Z_j,
\end{eqnarray} 
where $Z_j = Z_j^2=Z_j^\dag \in \H_+^k$ and $\sum_{j=1}^{n_z} Z_j = I$. 
\cite{hu2022robust} used \emph{facial reduction} (\cite{borwein1981facial}) for calculating the QKD rate. Their approach restricts the feasible region of the problem into a smaller face and reduces the dimension of the matrices, which improves the performance of interior-point methods. 
Another simplification by \cite{hu2022robust} is using the special structure of $\mathcal Z$ to prove the following equation for every $\delta \succeq I$:
\begin{eqnarray} \label{eq:QRE-19}
\trace(\delta \ln(\mathcal Z(\delta)) = \trace(\mathcal Z(\delta) \ln(\mathcal Z(\delta)).
\end{eqnarray}
This can simplify the QRE function as the difference of two quantum entropy (QE) functions, which makes calculating the gradients and Hessians much easier. Using these techniques, \cite{hu2022robust} designed an interior-point algorithm specialized just for computing the QKD rate.

\subsection{Two-phase approach}
The analytic facial reduction techniques by \cite{hu2022robust} can be used with DDS 2.2 as well. Here, we propose a two-phase approach for finding the minimal face of the positive semidefinite cone for the feasible region of the problem. A rationale is that solving QRE optimization is much costlier than solving SDPs. If we can use a Phase-I SDP to find a better-conditioned formulation for the QRE optimization problem, the overall cost will be lower. Our Phase-I SDP is based on the following lemma:

\begin{lemma}
Consider the spectrahedron defined by the following constraints:
\begin{eqnarray} \label{eq:QRE-13}
&&\langle A_i, \rho \rangle = b_i, \ \ \ i=\{1,\ldots,m\}  \nonumber \\
&&\rho \succeq 0.
\end{eqnarray}
Let $y \in \R^m$ be such that
\begin{eqnarray} \label{eq:QRE-14}
&&Y := \sum_{i=1}^m y_i A_i  \succeq 0 \nonumber \\
&& y^\top b = 0.
\end{eqnarray}
Then, for every $\rho$ in the spectrahedron defined by \eqref{eq:QRE-13}, we have $ \rho Y  = 0$. 
\end{lemma}
Indeed, as in Section \ref{sec:2phase}, we want a maximum rank solution to the system \eqref{eq:QRE-14}. Even then, this will correspond to a single iteration of full facial reduction. 
Consider a $Y \in \S_+^n$ from the lemma that has $\bar n$ zero eigenvalues with spectral decomposition 
\[
Y=[U \ \ V] \Diag(\lambda_1, \ldots, \lambda_{n-\bar n}, 0, \ldots, 0) [U \ \ V]^\top.
\]
Then, $\rho Y=0$ and $\rho \succeq 0$ imply that $\rho = V \bar \rho V^\top$ for some $\bar \rho \in \S_+^{\bar n}$, and the feasible region can be equivalently written as:
\begin{eqnarray} \label{eq:QRE-15}
&&\langle V^\top A_i V, \bar \rho \rangle = b_i, \ \ \ i\in \{1,\ldots,m\}  \nonumber \\
&&\bar \rho \succeq 0,
\end{eqnarray}
where the size of the $\bar \rho$ matrix was reduced to $\bar n$. For Phase-I of the optimization process (Dual method of Section \ref{sec:2phase}), we can approximately solve the following problem:

\begin{eqnarray} \label{eq:QRE-16}
& \inf &   -\ln(\det(Y))  \nonumber \\
&&Y := \sum_{i=1}^m y_i A_i  \succeq 0 \nonumber \\
&& y^\top b = 0.
\end{eqnarray}
The effect of using phase-I on the three groups of the QKD problems are shown in Table \ref{tab:1}. The main bottleneck in the speed of the code is the dimension of $\mathcal G(\rho)$ and $ \mathcal Z(\mathcal G(\rho))$ as the arguments of $qre$. We can significantly reduce this dimension by the following lemma:
\begin{lemma} \label{lem1}
Consider the Kraus operator $\mathcal G$ and assume $\mathcal Z (\mathcal G(I)) $ has $\bar n$ non-zero eigenvalues with the spectral decomposition 
\begin{eqnarray} \label{eq:QRE-17}
\mathcal Z (\mathcal G(I)) = \sum_{i=1}^{n_g} \sum_{j=1}^{n_z} Z_i K_j K_j^\dag Z_i^\dag = [U \ \ V] \Diag(\lambda_1, \ldots, \lambda_{\bar n}, 0, \ldots, 0) [U \ \ V]^\dag.
\end{eqnarray}
Then, we have 
\begin{eqnarray} \label{eq:QRE-18}
qre(\mathcal G(\rho), \mathcal Z(\mathcal G(\rho))) &=& qre( U^\dag \mathcal G(\rho) U,  U^\dag \mathcal Z(\mathcal G(\rho)) U).
\end{eqnarray}
\end{lemma}
\proof{Proof.}
Note that using \eqref{eq:QRE-19}, we have
\begin{eqnarray}
qre(\mathcal G(\rho), \mathcal Z(\mathcal G(\rho))) &=& \trace{G(\rho) \ln(G(\rho))} - \trace \mathcal Z(\mathcal G(\rho)) \ln(\mathcal Z(\mathcal G(\rho))) \nonumber \\
                   &=& \trace(F(\mathcal Z (\mathcal G(\rho)))) - \trace(F(U^\dag \mathcal Z(\mathcal G(\rho)) U)),
\end{eqnarray}
where $F$ is the matrix extension of $f(x):=x\ln(x)$. Therefore, to show \eqref{eq:QRE-18}, it suffices to show that $U^\dag \mathcal G(\rho) U$ has the same non-zero eigenvalues as $\mathcal G(\rho)$ and $U^\dag \mathcal Z(\mathcal G(\rho)) U$ has the same non-zero eigenvalues as $\mathcal Z(\mathcal G(\rho))$. Consider the columns of $V = [v_1 \ \ \ldots \ \ v_{n-\bar n}]$. We claim that 
\begin{eqnarray} \label{eq:QRE-20}
&& K_j^\dag Z_i^\dag v_t = 0, \ \ \ j\in\{1,\ldots, n_g\}, i \in \{1,\ldots,n_z\}, t\in\{1,\ldots,n-\bar n\}  \nonumber \\
&& K_j^\dag v_t = 0, \ \ \ j\in\{1,\ldots, n_g\}, t\in\{1,\ldots,n-\bar n\}.
\end{eqnarray}
For the first equation, note that for each $t\in\{1,\ldots,n-\bar n\} $ we have
\begin{eqnarray} \label{eq:QRE-21}
 0 &=& v_t^ \dag \mathcal Z (\mathcal G(I)) v_t  = \sum_{i=1}^{n_g} \sum_{j=1}^{n_z} v_t ^\dag Z_i K_j K_j^\dag Z_i^\dag v_t \nonumber \\
   &=& \sum_{i=1}^{n_g} \sum_{j=1}^{n_z} \|K_j^\dag Z_i^\dag v_t\|^2,
\end{eqnarray}
which implies the first equation in \eqref{eq:QRE-20}. For the second equation, we can use the first equation and the fact that $\sum_{i=1}^{n_z} Z_i = I$: For each fixed $j$ and $t$, we can add the the equations for all $i \in \{1,\ldots,n_z\}$. Equation \eqref{eq:QRE-20} is important since it shows that for any $\rho$, $\mathcal G(\rho)$ and $\mathcal Z(\mathcal G(\rho))$ have the columns of $V$ in their null space. Therefore, the range of $U$ contains the ranges of $\mathcal G(\rho)$ and $\mathcal Z(\mathcal G(\rho))$ for any matrix $\rho$. This implies the statement of the lemma. Specifically, if $\gamma$ is a non-zero eigenvalue of $\mathcal Z(\mathcal G(\rho))$ with eigenvector $w$, there exists $\bar w$ such that $w = U \bar w$, then
\[
U^\dag \mathcal Z(\mathcal G(\rho)) U \bar w = U^\dag \mathcal Z(\mathcal G(\rho)) w = U^\dag (\gamma w) = \gamma U^\dag w. 
\]
  \endproof
In Subsection \ref{subsec:QKD}, we use the methods we developed here to calculate the QKD rate for multiple protocols. 
\section{Handling Complex Matrices} \label{sec:4} 
For some applications, including key rate calculations for QKD channels, we need to handle complex Hermitian matrices. For software packages such as DDS that only accept real symmetric matrices, we need an equivalent formula for $qre(X,Y)$ based on the real and imaginary parts of $X$ and $Y$. For a unitary matrix $U = U_r + \iota U_i$ (where $\iota=\sqrt{-1}$), we have
\begin{eqnarray} \label{eq:un-1}
UU^*=U^*U=I  \Leftrightarrow  \left\{\begin{array}{l}
U_rU_r^\top + U_iU_i^\top = I \\
-U_rU_i^\top + U_iU_r^\top = 0
\end{array} \right..
\end{eqnarray}
For any complex $n$-by-$n$ matrix $X=X_r + \iota X_i$, we define a $2n$-by-$2n$ matrix $\bar X$ as
\[
\bar X = \left[\begin{array}{cc}
X_r & -X_i \\  X_i & X_r
\end{array} \right].
\] 
If $X$ is Hermitian, then $\bar X$ is symmetric. 
By using \eqref{eq:un-1}, we can show that if $U$ is a unitary matrix, then $\bar U$ is also a real unitary matrix. We can also show that if $X=UDU^*$ is a spectral decomposition for $X$, then 
\[
\bar X = \bar U  \Diag(D,D) \bar U^\top,
\]
is a spectral decomposition for $\bar X$. This implies that for every function $f: \mathbb R \rightarrow \mathbb R\cup\{+\infty\}$ we have
\[
\trace(F(\bar X)) = 2 \trace(F(X)).
\]
Now we can prove the following lemma:
\begin{lemma}
For two Hermitian matrices $X,Y \in \H_+^n$ we have
\[
qre(\bar X, \bar Y) = 2 qre(X,Y). 
\]
\end{lemma}
\proof{Proof.}
First we show that for two Hermitian matrices $X,W \in \H_+^n$, we have
\begin{eqnarray} \label{eq:QRE-23}
\trace(\bar X \bar W) = 2 \trace(XW).
\end{eqnarray}
Assume that $X=X_r+\iota X_i$ and $W=W_r+\iota W_i$. Then, we have
\[
\trace(XY) = \trace(X_rW_r-X_iW_i+\iota (X_rW_i+X_iW_r)) = \trace(X_rW_r-X_iW_i),
\]
where for the last equation we used the fact that $\trace(XY)$ is a real number. Then, we have
\[
\trace(\bar X \bar Y) = \trace\left( \left[\begin{array}{cc}
X_r & -X_i \\  X_i & X_r
\end{array} \right] \left[\begin{array}{cc}
W_r & -W_i \\  W_i & W_r 
\end{array} \right]\right)= 2\trace(X_rW_r-X_iW_i).
\]
The last two equations confirm \eqref{eq:QRE-23}. To complete the proof, we claim that for every function $F$, we have
\begin{eqnarray} \label{eq:QRE-24}
\overline {F(X)} = F(\bar X).	
\end{eqnarray}
Assume that the spectral decomposition of $X$ is $X=UDU^*$. 
\begin{eqnarray*}
F(X)=U F(D) U^*&=&(U_r+\iota U_i) F(D) (U_r^\top-\iota U_i^\top)  \\
               &=& U_rF(D)U_r^\top+U_iF(D)U_i^\top + \iota (-U_rF(D)U_i^\top+U_iF(D)U_r^\top).
\end{eqnarray*}
Therefore, we have
\begin{eqnarray*}
\overline {F(X)} = \left[\begin{array}{cc}
U_rF(D)U_r^\top+U_iF(D)U_i^\top & U_rF(D)U_i^\top-U_iF(D)U_r^\top \\  -U_rF(D)U_i^\top+U_iF(D)U_r^\top & U_rF(D)U_r^\top+U_iF(D)U_i^\top
\end{array} \right].
\end{eqnarray*}
The spectral decomposition of $\bar X$ is $\bar X = \bar U \Diag(D,D) \bar U^\top$. Therefore, \\
$F(\bar X)=\bar U \Diag(F(D),F(D)) \bar U^\top$. By expanding this term, we can confirm that  \eqref{eq:QRE-24} holds. Now, \eqref{eq:QRE-23} and \eqref{eq:QRE-24} imply the result of the lemma. If we define $F(X):=\ln(X)$, then
\begin{eqnarray*}
\begin{array}{cclr}
qre(\bar X, \bar Y) & = & \trace(\bar XF(\bar X)) - \trace(\bar XF(\bar Y)) & \\
                    & = & \trace(\bar X \overline {F( X)}) - \trace(\bar X \overline {F(Y)}), &   \ \ \ \text{by using }\eqref{eq:QRE-24}\\
                    & = & 2 \trace{XF(X)} - 2 \trace(XF(Y)),  & \ \ \ \text{by using } \eqref{eq:QRE-23}  \\
                    & = & 2 qre(X,Y).
 \end{array}
\end{eqnarray*} 
  \endproof
\section{Numerical Results} \label{sec:num} 
The techniques designed here have been used to improve the performance of the newest version of the software package DDS (namely DDS 2.2), which can be downloaded from \cite{DDS2.2}.

DDS accepts every combination of the following function/set constraints: (1)  symmetric cones (LP, SOCP, and SDP); (2) quadratic constraints that are SOCP representable; (3) direct sums of an
arbitrary collection of 2-dimensional convex sets defined as the epigraphs of univariate convex
functions (including as special cases geometric programming and entropy programming); (4) generalized  Koecher (power) cone; (5) epigraphs of matrix norms (including
as a special case minimization of nuclear norm over a linear subspace); (6) vector relative entropy; (7) epigraphs of quantum
entropy and quantum relative entropy; and (8) constraints involving hyperbolic polynomials.  The command in MATLAB that calls DDS has the following form (see \cite{DDS2.2})
\begin{eqnarray} \label{eq:dds-form}
\tx{[x,y,info]=DDS(c,A,b,cons,OPTIONS)}.
\end{eqnarray}

\noindent {\bf Input Arguments:} \\
\noindent \tx{cons}:  A cell array that contains the information about the type of constraints. \\
\noindent \tx{c,A,b}: Input data for DDS: $A$ is the coefficient matrix, $c$ is the objective vector, $b$ is the shift vector. \\
\noindent \tx{OPTIONS} (optional): An array which contains information about the tolerance  and initial points.  

\noindent {\bf Output Arguments:} \\
\noindent \tx{x}: Primal point. \\
\noindent \tx{y}: Dual point which is a cell array. \\
\noindent \tx{info}: A structure array containing performance information such as \tx{info.time}, which returns the CPU time for solving the problem.

In this section, we present several numerical examples of running DDS 2.2 for QRE programming. We performed computational experiments using the software MATLAB R2022a, on a 1.7 GHz 12th Gen Intel Core i7 personal computer with 32GB of memory. All the numerical results in this section are by using the default settings of DDS, including the tolerance of $tol=10^{-8}$. Some of the examples in this section are included in a developing work by the authors to create a library for multiple classes of modern convex optimization problems (\cite{KarimiLib}). 
\subsection{Nearest correlation matrix}
The nearest correlation matrix in the classical sense has been heavily studied for years, for example by \cite{higham2002computing,HSS2016}. Here, we are interested in the nearest correlation matrix in the \emph{quantum sense}, which was introduced in the example folder of CVXQUAD (\cite{cvxquad}) and then adopted by Hypatia (\cite{coey2022solving}) for numerical experiments.
For a fixed matrix $M \in \S^n_+$, the nearest correlation matrix in the quantum sense is defined as a matrix $Y_M$ with all diagonals equal to 1 that minimizes $qre(M,Y)$. In other words:
\begin{eqnarray} \label{eq:QRE-22}
\begin{array}{cll}
Y_M = \textup{argmin} &  qre(M,Y) & \\
 & Y_{ii} = 1, & i\in\{1,\ldots, n\}.
\end{array}
\end{eqnarray}
To make a comparison  with Hypatia which uses the exact formulation for the Hessian, we consider a fixed matrix $M \in \S^n$ and change $n$ to see how DDS 2.2 and Hypatia scale in this problem. For the numerical experiments, we assume that $Y$ is a tridiagonal matrix with all the diagonals equal to one. We consider two cases for $M$: 1) $M=2I$, and 2) $M$ is a random positive definite matrix constructed as
 \begin{eqnarray} \label{eq:higham-2}
M :=  M_0M_0^\top  / \|\diag(M_0M_0^\top)\|_\infty,
\end{eqnarray}
where $M_0:=\tx{rand(n)}$ is an $n$-by-$n$ matrix of uniformly distributed random numbers over $(0,1)$. 
Table \ref{tab:DDS-Hy} shows the iterations and time that both DDS 2.2 and Hypatia take to solve the problem for different values of $n$. For the random $M$ cases, the results are averaged over 10 instances. 
\begin{table} [!ht] 
 \caption{Finding the nearest correlation matrix in the quantum sense using DDS 2.2 and Hypatia. Times are in seconds. \label{tab:DDS-Hy}}
  \centering
  \renewcommand*{\arraystretch}{1.1}
  \begin{tabular}{ |c | c| c | c| c| }
    \hline
 & \multicolumn{2}{|c|}{$M=2I$}  &  \multicolumn{2}{|c|}{Random $M$ (average)} \\   \hline 
n  &    DDS    Itr/time    & Hypatia Itr/time & DDS    Itr/time    & Hypatia Itr/time\\  \hline
 25 &    11/\ 0.8   &     15 /\ 6.6            & 30/\  5 & 19/\ 4\\ \hline 
   50 &   14/\ 3.6    &    15/\ 43             & 37/\  20 & 27 /\ 29 \\ \hline 
   75 &   17/\ 13.7    &      17/\ 248         & 51 /\ 65 & 33 /\ 152 \\ \hline 
    100 &   19/\ 32    &    18/\ 900           & 58 /\ 168 & 40 /\ 829  \\ \hline 
 125 &   21/\ 50.3    &      20/\ 1993         & 62 /\ 375 & 43 /\ 2701 \\ \hline 
  150  &  22/\ 92.53  &   20 /\ 5627           & 66 /\ 693 &  46 /\ 6079 \\ \hline
  175 &   24/\ 139.8    &   time $> 10^4$      & 71 /\ 1184 & time $> 10^4$\\ \hline 
  200 &   26/\ 237    &   time $> 10^4$        & 75 /\ 1760 & time $> 10^4$\\ \hline 
  250 &   30/\ 550    &   time $> 10^4$        & 78 /\ 3501 & time $> 10^4$\\ \hline 
  300 &   32/\ 1080  & time  $>10^4$           & 80 /\ 6980 & time $> 10^4$\\ \hline 
\end{tabular}
\end{table}
As can be seen, the running time explodes fast by increasing the dimension of the matrices for Hypatia, where for DDS 2.2, the increase rate is more reasonable due to the techniques used in the paper. For more numerical examples, we use the matrices introduced in \cite{HSS2016} for classical correlation matrix problem defined as
\begin{eqnarray} \label{eq:higham-1}
M_0:= \left [\begin{array}{cc}
A  & Y \\ Y^\top & B
\end{array}  \right],
\end{eqnarray}
where $A \in \S^m_{++}$ is a random correlation matrix, $B \in \S^n$ is a random matrix with uniformly distributed numbers and all 1 diagonal, and $Y$ is an $m$-by-$n$ random matrix with uniformly distributed numbers. Since for the quantum nearest correlation matrix $M$ needs to be positive definite, we define $M$ using $M_0$ and equation \eqref{eq:higham-2}.

Table \ref{tab:DDS-Hy-2} shows the iterations and time that both DDS 2.2 and Hypatia take to solve the problem for different values of $(m,n)$. Note that for the largest instances reported in Table \ref{tab:DDS-Hy-2}, the arguments of the $qre$ function are 400-by-400 matrices. 
\begin{table} [!ht] 
 \caption{Finding the nearest correlation matrix in the quantum sense for $M$ defined by \eqref{eq:higham-1} and \eqref{eq:higham-2} using DDS 2.2 and Hypatia. Times are in seconds. \label{tab:DDS-Hy-2}}
  \centering
  \renewcommand*{\arraystretch}{1.1}
  \begin{tabular}{ |c | c| c |}
    \hline
(m,n)  &    DDS    Itr/time    & Hypatia Itr/time  \\ \hline
 (25,25)  &   31/\ 16   &     19 /\ 25         \\ \hline 
 (40,40)  &     37/\ 52.8     &   24/\ 437  \\ \hline
 (40,60)  &     40/\ 134     &   24/\ 980  \\ \hline
 (60,90)  &     49/\ 305     &   time  $>10^4$  \\ \hline
 (100,100)  &     57/\ 1033     &   time  $>10^4$  \\ \hline
 (200,100)  &     64/\ 4111     &   time  $>10^4$  \\ \hline
 (200,200)  &     72/\ 9660     &   time  $>10^4$  \\ \hline

\end{tabular}
\end{table}

\subsection{QRE with other type of convex constraints}
DDS is a software package for convex optimization which accepts a combination of multiple conic and non-conic constraints (\cite{DDS2.2}). Considering QRE programming, DDS lets us solve problems with QRE constraints combined with several other constraints. As far as we know, DDS is the only available software that can solve QRE problems of these sizes combined with other types of constraints. Moreover, DDS is the only available code to handle some types of constraints such as the ones involving hyperbolic polynomials. Preliminary results of QRE programming was reported for DDS 2.1 (\cite{DDS}). 
To compare DDS 2.1 and 2.2 for QRE programming, we run the same table in (\cite{DDS}) for DDS 2.1 and re-run it for DDS 2.2 (\cite{DDS2.2}). The results are given in Table \ref{tab:DDS}. We also added the results of using CVXQUAD created by \cite{fawzi2019semidefinite}, which uses SDP approximation for the QRE programming. As can be seen, CVXQUAD does not scale well, and we have size error even for problem instances with 20-by-20 matrices. 
\begin{table} [ht] 
 \caption{Results for problems involving Quantum Relative Entropy using DDS 2.1, DDS 2.2, and CVXQUAD (with SDPT3 as the solver) \label{tab:DDS}}
 \centering 
  \renewcommand*{\arraystretch}{1.1}
  \begin{tabular}{ |l | c| c | c | c |  c| c|}
    \hline
Problem  &    size of $A$     &    Itr/time(sec)  & Itr/time(sec) & Itr/time(sec)\\
 &       &         DDS 2.1  &  DDS 2.2 & CVXQUAD \\ \hline 
QuanReEntr-6  & $73*13$  &  9/\ 1.0  &  12/ \ 0.8 & 20/\ 1.7  \\ \hline
QuanReEntr-10  & $201*21$  & 12/\  11.2  &  12/ \ 1.2  & 25/\ 95\\ \hline
QuanReEntr-20  & $801*41$  &  15/\  34.4  &  15/\ 1.2  & Size error\\ \hline
QuanReEntr-LP-6  & $79*13$  &  29/\ 1.7  &  25/ \ 0.7  & 24/\ 4.8\\ \hline
QuanReEntr-LP-6-infea  & $79*13$  &  30/\ 1.7  &  28/ \ 0.8 & 28/\ 3.8 \\ \hline
QuanReEntr-LP-10  & $101*21$  &  27/\ 4.6  &  27/ \ 1.3 &  38 /\ 678.9\\ \hline
\end{tabular}
\end{table}
Due to the major improvements in DDS 2.2 which we are reporting here, we can now solve much larger instances. Consider an optimization problem of the form
\begin{eqnarray} \label{eq:QRE-25}
\begin{array}{ccll}
\textup{min} &  & qre\left(A_0+\sum_{i=1}^k x_i A_i, B_0+\sum_{i=1}^k x_i B_i\right) & \\
 & (I)& x \geq b_L, &  \\
 & (II) &\| x - b_N \|_p  \leq \alpha, \\
 & (III) & p(x+b_H) \geq 0.
\end{array}
\end{eqnarray}
For the created examples, $A_i$ and $B_i$, $i\in\{1,\ldots,k\}$, are sparse 0,1 random symmetric matrices. 
Table \ref{tab:DDS2} shows the results of running DDS 2.2 on some instances of the form \eqref{eq:QRE-25}. QRE-LP problems only have (I) as the constraint. QRE-LP-POW3 and QRE-LP-SOCP problems  have (I)-(II) as constraints, with respectively $p=3$ and $p=2$. The problems with infeas in the name are infeasible. Exploiting duality in an efficient way makes DDS robust in detecting the infeasibility of the problems (\cite{karimi_status,DDS}). The problems QRE-Vamos has (III) as the constraint where $p$ is a hyperbolic polynomial created by Vamos-like matroids as explained in \cite{DDS}. QRE-KL problems are of the form:
\begin{eqnarray} \label{eq:QRE-26}
\begin{array}{ccll}
\textup{min} &  & qre\left(A_0+\sum_{i=1}^k x_i A_i, B_0+\sum_{i=1}^k y_i B_i\right) & \\
 & & KL(x,y)  \leq \gamma, &  \\
\end{array}
\end{eqnarray}
where $KL$ is defined in \eqref{eq:KL-1}. 
\begin{table} [!ht] 
 \caption{Results for problems involving Quantum Relative Entropy using DDS 2.2. The types of the constraints are included in the name of the problem. The number is the size of the matrices in the QRE constraint. Vamos stands for hyperbolic polynomials created by Vamos-like matroids.  \label{tab:DDS2}}
  \centering
  \renewcommand*{\arraystretch}{1.1}
  \begin{tabular}{ |l | c| c | c | c |  c| }
    \hline
Problem  &    size of \tx{cell2mat(A)}     &    Itr/time(sec)   \\
 &      in \eqref{eq:dds-form}  &         DDS 2.2  \\ \hline 
QRE-LP-100  & $20101\times 101$  &  17/\ 42    \\ \hline
QRE-LP-200-infeas  & $80201\times 201$  &  50/\ 966    \\ \hline
QRE-LP-Pow3-20 & $842\times 21$  &  37/\ 6.7    \\ \hline
QRE-LP-Pow3-20-infeas & $842\times 21$  &  25/\ 3.8    \\ \hline
QRE-LP-Pow3-100 & $20201\times 101$  &  56/\ 170  \\ \hline
QRE-LP-Pow3-100-infeas & $20201\times 101$  &  44/\ 74  \\ \hline
QRE-LP-SOCP-20& $842\times 21$  &  39/\ 4.2  \\ \hline
QRE-LP-SOCP-20-infeas & $842\times 21$  &  24/\ 3.9  \\ \hline
QRE-LP-SOCP-100 & $20202\times 101$  &  66/\ 180  \\ \hline
QRE-LP-SOCP-100-infeas & $20202\times 101$  &  28/\ 90  \\ \hline
QRE-LP-SOCP-200-infeas & $80402\times 201$  &  36/\ 544  \\ \hline
QRE-LP-SOCP-200 & $80402\times 201$  &  103/\ 1191  \\ \hline
QRE-Vamos-20 & $811\times 6$  &  22/\ 3.6  \\ \hline
QRE-Vamos-20-infeas & $811\times 6$  &  32/\ 7.8  \\ \hline
QRE-Vamos1-100 & $20011\times 6$  &  27/\ 48  \\ \hline
QRE-Vamos2-100 & $20011\times 6$  &  44/\ 129  \\ \hline
QRE-KL-100  & $20202 \times 201$  & 49/\ 109  \\ \hline 
QRE-KL-100-infeas  & $20202 \times 201$  & 22/\ 37  \\ \hline 
QRE-KL-200  &  $80402 \times 401$ & 76/\ 1153  \\ \hline
QRE-KL-200-infeas  &  $80402 \times 401$ & 25/\ 278  \\ \hline

\end{tabular}
\end{table}

\subsection{Two-phase methods for QRE programming} \label{subsec:2p}
We proposed two-phase methods for QRE programming in Section \ref{sec:2phase}. For numerical experiments, we have synthesized some problems by fixing $r$ and $n$, and choosing a $n$-by-$r$ matrix $V$ which is all zero except the main diagonal is all ones. Let $E_i \in \S^r$ be a matrix of all zeros except a 1 on the $i$th diagonal entry. Consider problem \eqref{eq:QRE-28} where $M:=I$ and the linear constraints are $x_i \leq 1$ for $i\in \{1,\ldots,k\}$. We define
\[
A_i := VE_iV^\top,  \ \ i \in \{1,\ldots,k\}. 
\]
Table \ref{tab:2phase} shows the results of solving problem \eqref{eq:QRE-28} using DDS 2.2 for different values of $r$ and $n$. We explained that problems \eqref{eq:QRE-29} and \eqref{eq:QRE-28-2} are minimizing a s.c.\ barrier which can be done very efficiently. Since the current version of DDS does not support this, we use the SDP reformulations for Phase-I. As can be seen, the reformulated QRE after phase-I is not only smaller in size, but the much fewer number of iterations shows that it is more well-conditioned. The overall running time of the two-phase method is smaller than the 1-phase one, and the gap grows by increasing $n$.  Note that both Phase-I SDP and Phase-II QRE programming are solved by using DDS. 

\begin{table}[h!t]
\caption{The effect of the primal and dual two-phase approachs on the running time and condition of the QRE problem. For the primal two-phase method, a feasible initial point is also given as input to DDS.    \label{tab:2phase}}
\centering
\begin{tabular}{|c|c|c|c|c|c|c|}
\hline
n& r  & \multicolumn{2}{|c|}{Dual two-phase \eqref{eq:QRE-29}}  & \multicolumn{2}{|c|}{Primal two-phase \eqref{eq:QRE-28-2}}& 1-phase method \\ 
                                 &                    & Phase-I time & iter/time   & Phase-I time & iter/time & iter/time \\ \hline

25 & 5 &  0.96 &  13 /\ 0.39 &0.25& 12/\ 0.35 & 90/\ 5.1 \\ \hline
50 & 5 &  0.91 &  12 /\ 0.45 &0.3&11 /\ 0.3& 108/\ 29 \\ \hline
100 & 5 &  5.8 &  12 /\ 0.42 &0.4 &11 /\ 0.4& 172/\ 178 \\ \hline
200 & 5 &  70 &  13 /\ 0.44 &0.7&10/\ 0.3& 226/\ 1527 \\ \hline
300 & 5 & 654  &  12/\ 0.74 & 2.9 & 11/\ 0.44    & 313/\ 8560 \\ \hline
25 & 10 &  1.1 &  14 /\ 0.6 &0.3&14/\ 0.4& 78/\ 4.0 \\ \hline
50 & 10 &  0.9 &  14 /\ 0.63 & 0.38 & 13/0.53& 97/\ 26.6 \\ \hline
100 & 10 &  5.3 &  15 /\ 0.68 & 0.49 & 13/0.55& 164/\ 180 \\ \hline
200 & 10 &  72 &  15 /\ 0.68 &0.78 & 14/\ 0.49 & 214/\ 1695 \\ \hline
300 & 10 &  736 &  15 /\ 0.72 &4.5 & 13/\ 0.57 & 304/\ 8663 \\ \hline
25 & 20 &  0.7 &  12 /\ 1 & 0.3&11/\ 0.9& 48/\ 3.2 \\ \hline
50 & 45 &  0.87 &  15 /\ 4.3 &0.32 &13/\ 3.4& 50/\ 14 \\ \hline
100 & 95 &  4.5 &  25 /\ 46 & 0.56 &16/\ 23& 54/\ 81.4 \\ \hline
200 & 195 &  63 &  32 /\ 342 &1.07 &23/\ 188& 50/\ 478 \\ \hline
300 & 295 &  621 &  38 /\ 1620 &2.04 &28/\ 785& 52/\ 1412 \\ \hline
\end{tabular}
\end{table}

\subsection{Symmetric QRE} \label{subsec:SQRE}
Handing symmetric QRE constraints using QRE ones discussed in Section \ref{sec:SQRE}. Consider the following two optimization problems:
\begin{eqnarray*} \label{eq:QRE-27}
\begin{array}{cll}
\textup{min} &   qre\left(I+\sum_{i=1}^k x_i A_i, I+\sum_{i=1}^k x_i B_i\right) & \\
 &  x \geq \ell, & 
\end{array}  \ \ \ 
\begin{array}{cll}
\textup{min} &   sqre\left(I+\sum_{i=1}^k x_i A_i, I+\sum_{i=1}^k x_i B_i\right) & \\
 &  x \geq \ell, & 
\end{array} 
\end{eqnarray*}
where $A_i \in \S^n$ and $B_i \in \S^n$, $i\in\{1,\ldots,k\}$, are sparse 0-1 random matrices (each matrix is all zero except for two off-diagonal entries). $x=0$ is a feasible solution for both problems. We want to compare the number of iterations and running time of solving these problems by changing $n$, the size of matrices where we choose $k=n$. Let us choose $\ell=-2 \mathbbm{1}$, where $ \mathbbm{1}$ is the vector of all ones. Table \ref{tab:SQRE} shows the results of the iteration count and running time for both problems using DDS 2.2. As can be seen in Table \ref{tab:SQRE}, the ratio of iteration counts are growing from 1 to 1.53, while the ratio of running times are growing from 1.7 to 2.45.  

\begin{table} [ht] 
 \caption{Comparing SQRE and QRE using DDS 2.2 \label{tab:SQRE}}
\centering  
  \renewcommand*{\arraystretch}{1.1}
  \begin{tabular}{ |l | c| c|}
    \hline
$n$  &       Itr/time(sec)  & Itr/time(sec) \\
  &          QRE &  SQRE  \\ \hline 
  10 &         9 /\ 0.3 &  9 /\ 0.5 \\ \hline 
  25 &         12 /\ 1.7 &  15 /\ 3.3 \\ \hline 
  50 &         13 /\ 4.6 &  13 /\ 8.1 \\ \hline
  100 &         14 /\ 28 &  15 /\ 47 \\ \hline
  150 &         14 /\ 81 &  16 /\ 159 \\ \hline
  200 &         14 /\ 166 &  17 /\ 333 \\ \hline
  250 &         15 /\ 344 &  23 /\ 858 \\ \hline
  300 &         15 /\ 698 &  23 /\ 1712 \\ \hline

\end{tabular}
\end{table}

\subsection{Quantum key distribution rate} \label{subsec:QKD}
We developed a two-phase approach for calculating the QKD rate in Section \ref{sec:QKD}. In this section, we consider some QKD protocols such as some variants of the Bennett-Brassard 1984 (BB84) protocol (\cite{bennett2014quantum}): entanglement-based (ebBB84),
prepare-and-measure (pmBB84), and measurement-device-independent (mdiBB84) (\cite{lo2012measurement}).  OpenQKDSecurity is a platform for  numerical key rate calculation of QKD (\cite{QKD2}), where several examples for different regimes can be created. One protocol for QKD is Entanglement-Based BB84. The parameters of the problem $\mathcal G$, $\mathcal Z$, and $\Gamma$ are calculated based on the probability of performing measurement in one of the two possible basis $p_z$, and the error rate $e$. As we use natural logarithm in DDS, the key rate $R$ for this protocol is calculated by the formula 
\[
R = \frac{p}{\ln(2)} - \delta_{EC},
\] 
where $p$ is the optimal value of \eqref{eq:num-3} and $\delta_{EC}$ is a constant caused by performing error-correction.

By applying the phase-I optimization discussed above and Lemma \ref{lem1}, not only do we significantly reduce the size of the involved matrices, but we also improve the condition of the problem.  Some examples are shown in Table \ref{tab:1}. Without using Phase-I, the problem is ill-conditioned and DDS cannot achieve the desired accuracy. 

\begin{table}[h!t]
\caption{The effect of phase-I and \ref{lem1} on reducing the size of $\rho$  \label{tab:1}}
\centering
\begin{tabular}{|c|c|c|c|c|c|}
\hline
protocol \ \  $(p_z,e)$ & $(n,k)$  & $(\bar n, \bar k)$ & \multicolumn{2}{|c|}{Phase-I and Lemma \ref{lem1}}  & Just Lemma  \ref{lem1} \\ 
                        &          &                    & Phase-I time & iter/time  & iter/time \\ \hline
pmBB84 \ \ $(0.5,.09)$  &  (32,8)  & (8,4) & 1 & 19/ 2.7   & 81/14\\ \hline
pmBB84 \ \ $(0.9,.09)$  &  (32,8)  & (8,4) & 0.9 & 19/ 2.7   & \text{ill-conditioned}\\ \hline
mdiBB84 \ \ $(0.5,.09)$  &  (96,48)  & (8,12)& 1.4  & 25/4.7 & \text{ill-conditioned}  \\ \hline
mdiBB84 \ \ $(0.9,.09)$  &  (96,48)  & (8,12)& 1.4  & 25/4.2 & \text{ill-conditioned}  \\ \hline
\end{tabular}
\end{table}

For the protocols eeBB84, pmBB84, and mdiBB84, the QRE program is setup by using two parameters: $p_z$ is the probability of choosing the $Z$ basis, and $e$ is the observed error rate. The iteration counts and running times of using DDS 2.2 for solving the QRE optimization problems for these three protocols are given in Tables \ref{table:QKD1}, \ref{table:QKD2}, and \ref{table:QKD3}. 

\begin{table} [!ht] 
  \caption{Numerical Report for ebBB84 Instances.}
  \label{table:QKD1}
  \centering
  \begin{tabular}{ |c | c| c | c | c|}
    \hline
Protocol & Parameters $(p_z,e)$ &    Size   &     Iter   &   Time  \\ \hline 
ebBB84  &  $(0.5,.01)$  &  (16,4)  & 23  & 0.8  \\  
ebBB84  &  $(0.5,.03)$  &  (16,4)  & 20  & 0.7  \\ 
ebBB84  &  $(0.5,.05)$  &  (16,4)  & 18  & 0.65  \\ 
ebBB84  &  $(0.5,.07)$  &  (16,4)  & 17  & 0.57  \\ 
ebBB84  &  $(0.5,.09)$  &  (16,4)  & 14  & 0.5  \\ 
ebBB84  &  $(0.7,.01)$  &  (16,4)  & 21  & 0.8  \\  
ebBB84  &  $(0.7,.03)$  &  (16,4)  & 21  & 0.74  \\ 
ebBB84  &  $(0.7,.05)$  &  (16,4)  & 17  & 0.65  \\ 
ebBB84  &  $(0.7,.07)$  &  (16,4)  & 16  & 0.6  \\ 
ebBB84  &  $(0.7,.09)$  &  (16,4)  & 17  & 0.65  \\ 
ebBB84  &  $(0.9,.01)$  &  (16,4)  & 22  & 0.8  \\  
ebBB84  &  $(0.9,.03)$  &  (16,4)  & 21  & 0.7  \\ 
ebBB84  &  $(0.9,.05)$  &  (16,4)  & 17  & 0.65  \\ 
ebBB84  &  $(0.9,.07)$  &  (16,4)  & 17  & 0.65  \\ 
ebBB84  &  $(0.9,.09)$  &  (16,4)  & 17  & 1.7  \\  \hline

\end{tabular}
\end{table}

\begin{table} [!ht] 
  \caption{Numerical Report for pmBB84 Instances.}
  \label{table:QKD2}
  \centering
  \begin{tabular}{ |c | c| c | c | c|}
    \hline
Protocol & Parameters $(p_z,e)$ &    Size   &     Iter   &   Time  \\ \hline 
pmBB84  &  $(0.5,.01)$  &  (32,8)  & 24  & 1.8  \\  
pmBB84  &  $(0.5,.03)$  &  (32,8)  & 22  & 1.4  \\ 
pmBB84  &  $(0.5,.05)$  &  (32,8)  & 21  & 1.3  \\ 
pmBB84  &  $(0.5,.07)$  &  (32,8)  & 18  & 1.3  \\ 
pmBB84  &  $(0.5,.09)$  &  (32,8)  & 17  & 1.1  \\ 
pmBB84  &  $(0.7,.01)$  &  (32,8)  & 24  & 1.6  \\  
pmBB84  &  $(0.7,.03)$  &  (32,8)  & 22  & 1.4  \\ 
pmBB84  &  $(0.7,.05)$  &  (32,8)  & 20  & 1.4  \\ 
pmBB84  &  $(0.7,.07)$  &  (32,8)  & 18  & 1.3  \\ 
pmBB84  &  $(0.7,.09)$  &  (32,8)  & 18  & 1.1  \\ 
pmBB84  &  $(0.9,.01)$  &  (32,8)  & 25  & 1.6  \\  
pmBB84  &  $(0.9,.03)$  &  (32,8)  & 23  & 1.5  \\ 
pmBB84  &  $(0.9,.05)$  &  (32,8)  & 21  & 1.1  \\ 
pmBB84  &  $(0.9,.07)$  &  (32,8)  & 20  & 1.1  \\ 
pmBB84  &  $(0.9,.09)$  &  (32,8)  & 21  & 1.4  \\  \hline

\end{tabular}
\end{table}  

\begin{table} [!ht] 
  \caption{Numerical Report for mdiBB84 Instances.}
  \label{table:QKD3}
  \centering
  \begin{tabular}{ |c | c| c | c | c|}
    \hline
Protocol & Parameters $(p_z,e)$ &    Size   &     Iter   &   Time  \\ \hline 
mdiBB84  &  $(0.5,.01)$  &  (96,48)  & 31  & 3.0  \\  
mdiBB84  &  $(0.5,.03)$  &  (96,48)  & 28  & 2.4  \\ 
mdiBB84  &  $(0.5,.05)$  &  (96,48)  & 28  & 3.1 \\ 
mdiBB84  &  $(0.5,.07)$  &  (96,48)  & 26  & 2.7  \\ 
mdiBB84  &  $(0.5,.09)$  &  (96,48)  & 18  & 1.6  \\ 
mdiBB84  &  $(0.7,.01)$  &  (96,48)  & 29  & 2.6  \\  
mdiBB84  &  $(0.7,.03)$  &  (96,48)  & 23  & 2.2  \\ 
mdiBB84  &  $(0.7,.05)$  &  (96,48)  & 24  & 2.4  \\ 
mdiBB84  &  $(0.7,.07)$  &  (96,48)  & 27  & 2.6  \\ 
mdiBB84  &  $(0.7,.09)$  &  (96,48)  & 26  & 2.6  \\ 
mdiBB84  &  $(0.9,.01)$  &  (96,48)  & 31  & 3.2  \\  
mdiBB84  &  $(0.9,.03)$  &  (96,48)  & 28  & 2.6  \\ 
mdiBB84  &  $(0.9,.05)$  &  (96,48)  & 24  & 2.3  \\ 
mdiBB84  &  $(0.9,.07)$  &  (96,48)  & 27  & 2.5  \\ 
mdiBB84  &  $(0.9,.09)$  &  (96,48)  & 26  & 2.7  \\  \hline

\end{tabular}
\end{table}

\section{Conclusion}
We developed novel numerical techniques to enhance the performance of interior-point (IP) methods which use the optimal self-concordant (s.c.)\ barrier function in \eqref{eq:QRE-bar}. Extensive numerical results demonstrate that DDS 2.2, which incorporates these techniques, can effectively solve significantly larger instances compared to its predecessor, DDS 2.1, and Hypatia.  The two-phase approaches proposed in this paper warrants further computational investigation in future research. The Phase-I problem can be tailored in various ways to modify the problem to help speed up Phase-II. Currently, the primary bottleneck in the speed of DDS 2.2 for QRE programming lies in calculating the matrix $S$ defined in \eqref{eq:QRE-8}. A significantly more efficient and numerically stable algorithm to implicitly or explicitly compute the matrix could significantly accelerate DDS and other IP solvers for QRE programming. Additionally, exploring the duality setup for the QRE cone presents an intriguing open question. A numerically robust characterization of the dual cone or the LF conjugate of the s.c.\ barrier could be leveraged in DDS and other solvers to further enhance their performance.

\renewcommand{\baselinestretch}{1}
\bibliographystyle{siam}
\bibliography{References}

\end{document}